# PARTNERSHIPS FOR PHYSICS TEACHING REFORM
## —a crucial role for universities and colleges

*David Hestenes and Jane Jackson*
Department of Physics and Astronomy
Arizona State University
Tempe, Arizona 85287-1504

**Abstract:** To meet *National Standards* recommended by the National Research Council for high school physics, inservice teachers must be integrated into the physics community. They must be *empowered* by *access* to resources of the physics community and by *sustained support* for their professional development. To that end, university and college physics departments must assume an essential role in establishing and maintaining the necessary *local infrastructure*. Nevertheless, this can be done within their current systems at negligible cost by forming *partnerships* with *local alliances* of inservice teachers. Practical details have been worked out, and working partnerships have been established in several localities. This effort should be extended to a national infrastructure for reform of physics teaching in community colleges as well as in high schools. Most teachers are eager to participate in teaching reform.

## 1. Introduction

Representing a broad consensus within the scientific and educational communities, the *National Research Council* (NRC) has recently published detailed *National Standards* for K-12 science education (1). That document is a call to action with guidelines for extensive reforms in science teaching and the professional development of teachers. University (and college) physics departments must play a role in the reform of high school physics, as the training of physics teachers has traditionally been their responsibility. However, this responsibility has been too narrowly conceived in the past and must be broadened if teaching reform is to be successful.

The *Standards* emphasize that "coherent and integrated programs" supporting "lifelong professional development" of science teachers are essential for significant reform. They state that "The conventional view of professional development for teachers needs to shift from technical training for specific skills to opportunities for intellectual professional growth." Such a program cannot be consistently maintained and enriched in any locality without dedicated support from a local university or college. Therefore, *physics departments must assume responsibility for supporting a program of sustained professional development for local inservice physics teachers.*

Incredibly, *this responsibility can be effectively discharged without significant costs to the university or new burdens on the physics faculty*. As explained below, inservice physics teachers are able and eager to take responsibility for their own professional development. All they need is support from faculty to offer well-designed professional development courses for university credit. Such courses are best taught by exceptional inservice teachers, so the demands on physics faculty are minimal. A local alliance of physics teachers can ensure that the courses are filled, so costs to the university are trivial or nil.

With professional development courses as an academic anchor, a partnership between university physics faculty and a local physics teacher alliance can achieve many other good things. For example, it provides an avenue for recruiting physics majors to the university and for effective outreach programs to the schools. However, to be optimally effective a partnership must have strong connections with current physics education research and curriculum development. We are currently running an NSF-funded project to supply such connections (2), and its unqualified success provides a "proof of concept" for the recommendations in this paper. However, the project needs to be expanded if it is to be available nationwide. That will require a broad commitment from the U.S. physics community.

State universities are best equipped to support partnerships for the professional development of inservice physics teachers, and their responsibility is the greater because they are supported by public funds. However, some private universities and colleges have concerned faculty and the necessary resources, so they should be encouraged to participate. Though we focus on high school physics teachers in this paper, community college teachers are equally in need of professional development. They should be encouraged to participate in local alliances with high school teachers, though they have some special needs of their own.

## 2. Empowering teachers—the key to reform

Judged by the National Standards (1) and a recent AIP survey (3), high school physics nationwide is badly in need of reform. The problems are many, but one thing is certain: Only the teachers can do it! The action is in the classroom. Physics cannot be "canned" in a "teacher-proof" textbook or curriculum materials. Physics is a way of thinking that requires skillful coaching by the teacher to develop.

Unfortunately, most physics teachers are poorly prepared for the task and lack the resources to improve. According to a recent nationwide AIP survey (3), about 70% believe that they are well-prepared to teach physics, though only 30% have a B.A. in physics or physics education. Only 15% feel qualified to discuss recent developments in physics. Even fewer are aware of advances in physics education research with practical implications for teaching. We estimate that less than 3% have implemented significant research-based reforms in their teaching.

Of the 18,000 physics teachers nationwide, 81% are the only physics teacher in the school. Despite being overworked, under-prepared, ill-informed and isolated from their peers, physics teachers have remarkably *strong élan and*

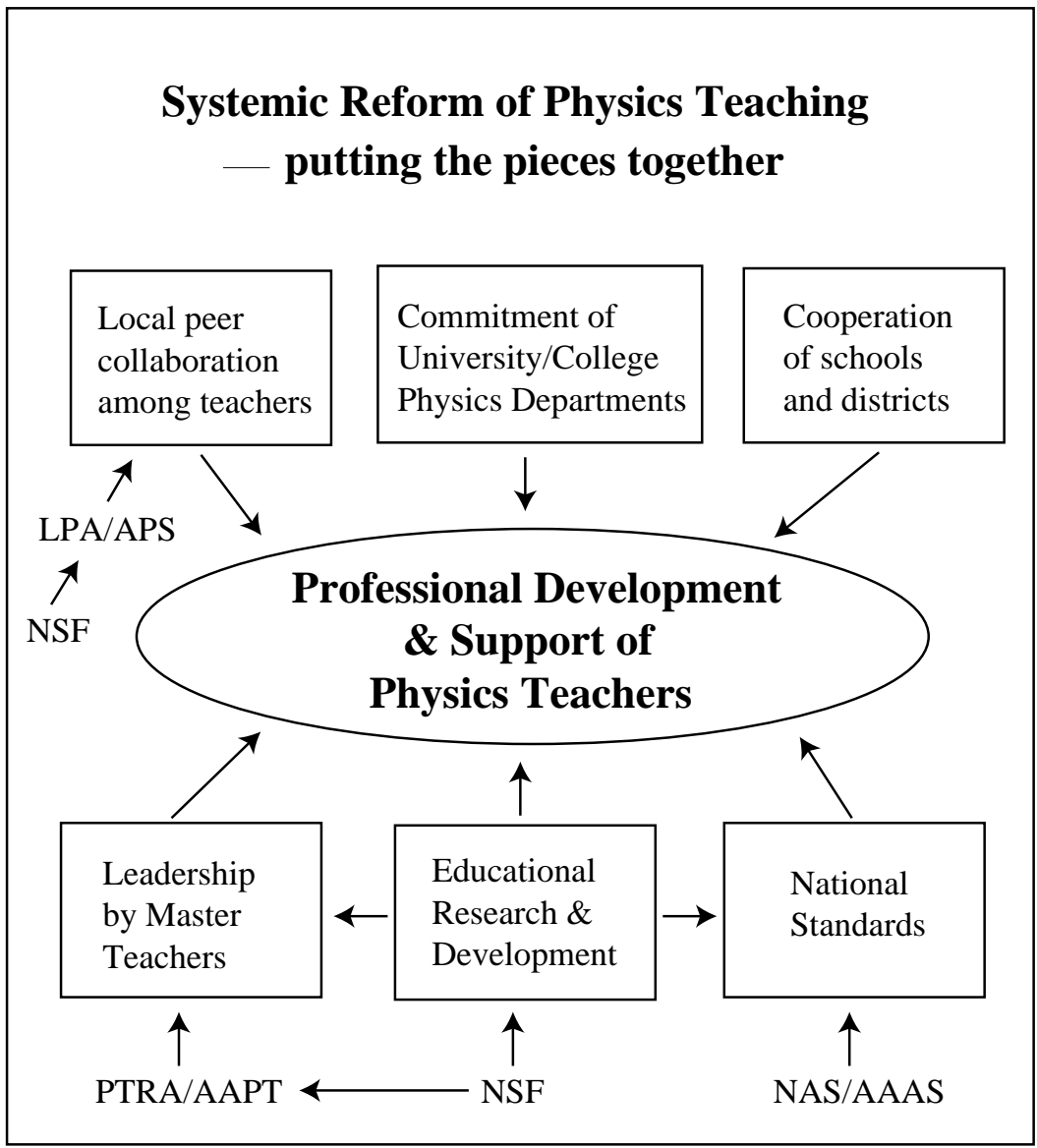

**FIGURE 1.**

*satisfaction* (3). More than 90% want to teach physics as often as possible, though only 35% teach it every year. In our program, so far involving teachers from 23 states (2), we have found that *most teachers are eager and able* to work hard and productively with their peers at improving their teaching. All they need is the opportunity and well-informed guidance. However, their efforts must be extended over many years to produce substantial improvements, so a stable infrastructure is needed to support it.

As the *National Standards* maintain (1): "The current reform effort in science education requires a substantive change in how science is taught. Implicit in this reform is an equally substantive change in professional development practices. . . . Professional development for teachers should be analogous to professional development for other professionals. Becoming an effective science teacher is a continuous process that stretches across the life of a teacher, from his or her undergraduate years to the end of a professional career. . . . Teachers require the opportunity to study and engage in research on science teaching and learning, and to share with colleagues what they have learned. . . . Practicing teachers have the opportunity to become sources of their own growth as well as supporters of the growth of others. Teachers should have opportunities for structured reflection on their teaching practice with colleagues, for collaborative curriculum planning, and for active participation in professional teaching and scientific networks. The challenge of professional development for teachers of science is to create optimal collaborative learning situations in which the best sources of expertise are linked with the experiences and current needs of the teachers."

To meet this challenge, physics teachers everywhere must be embraced by the physics community as respected professional colleagues. *Physics teachers are the representatives of the physics community in their classrooms.* If they are to convey the message that physics is our common cultural heritage — that physics belongs to everyone — if they are to wipe out the widespread perception of physics as an elitist foreign culture — if they are to promote access to physics for everyone, then they themselves must feel secure as respected members of the physics community. Token respect is not enough. Genuine respect by the physics community can be expressed only by fulfillment of its responsibility to support the professional development of teachers and the improvement of teaching conditions in the schools.

To establish and maintain a coherent program for *sustained* professional development and support of physics teachers will require leadership and commitment from the physics community. As indicated in Fig. 1, the pieces are in hand, many of them developed with NSF funding through the AAPT, APS, NAS, and the AAAS, but they are yet to be assembled. There are five major pieces:
- *National Standards* informed by expertise in science and educational research
- *Peer support* and collaboration among teachers
- *Leadership by master teachers* who are well-informed about educational research and curricula
- *Commitment from physics departments* in universities and colleges
- *Cooperation of schools,* school districts and communities.

Each piece will be described more fully before a plan to assemble them is proposed.

## 3. Peer support through Local Physics Alliances

As in any other profession, for continued professional growth teachers need the stimulus of interaction with their peers. Because so few have colleagues in their schools who share their interest in physics, physics teachers must look to professional organizations for peer contact. They hunger for such contact. Nearly half of them belong to the NSTA and/or the AAPT, but these organizations do not satisfy their need for regular peer contact. Recognizing this need, the APS *Local Physics Alliance* (LPA) program has organized physics teachers in high schools, colleges and universities across the country into local alliances to reduce the isolation and provide support for individual teachers. This is the beginning of a teacher support system that can be enhanced by electronic networking and professional development programs. University physics departments should take advantage of this powerful mechanism for reform of high school physics by supporting alliances in their localities and helping to organize new ones when necessary.

*Teacher ownership* and control of each local alliance is essential, because it is a professional organization by and for the teachers. The relation of the alliance to the university should be one of *partnership* with the common objective of promoting improvements in physics teaching. Identification of university faculty who are willing to support this partnership is essential. Through such *faculty partners* the alliances can draw on the resources of the university. This could include arrangements for a suitable meeting place for the alliance as well as information and advice about physics and physics teaching.

Internet electronic networking of all teachers in an alliance should be regarded as essential. Universities are ideally situated to help organize and support such networks. Universities can also provide valuable technical advice and support for computer infusion into high school physics.

There are now about 150 LPAs across the country, with memberships ranging from less than ten to well over a hundred. Surveys show that members value the LPA, reading the newsletter even if they do not attend meetings. The LPA has increased their awareness of student needs and the school situations of colleagues. However, many teachers are reluctant to become more involved because they cannot afford the time. This reluctance is likely to dissolve in LPAs which are of direct help to teachers in improving their teaching. That can be done!

Although some LPAs are very popular, none has come close to realizing its full potential, largely because they have a limited sense of purpose, but also because of meager support by the physics community. The avowed purpose of the LPAs is to share ideas about teaching, to learn about teaching strategies and new developments in physics. In reality, the LPAs contribute little to the improvement of teaching practice, and their main benefit is social. They could do better!

The LPAs lack systematic programs to improve physics pedagogy, and they have failed to recognize the power of organizing teachers into (action research) teams, which work on specific projects to improve teaching practice and report to their peers for critique and evaluation. In short, the *LPAs have not learned to function as a learning community to improve teaching*. To realize their potential they need informed leadership and partnerships with universities to establish coherent programs for professional development.

## 4. Programs for Professional Development—a role for universities

Most states encourage the professional development of teachers by coupling salary levels to academic credit from universities. Unfortunately, most post-baccalaureate coursework is so incoherent that it contributes little to the enhancement of teacher expertise. Universities can do much better by coordinating course offerings with the needs and desires of local teacher alliances — in other works, by offering the teachers greater control of their own professional development.

The most useful university course for high school physics teachers is a *Methods of Physics Teaching* course which thoroughly addresses *all* aspects of high school physics teaching, including the integration of teaching methods with course content as it should be done in the high school classroom. The course should incorporate up-to-date
- results of physics education research
- best high school curriculum materials
- use of technology
- experience in collaborative learning and guidance.

A two semester course is needed to cover the main topics of introductory physics.

In the past, those few universities which have offered such a course have intended it for training preservice physics teachers. No wonder they had trouble getting the enrollment to sustain the course — nationwide, only about 150 degrees for physics teaching are awarded each year! The enrollment problem disappears when the course is reconceived to meet the needs of inservice physics teachers — for the vast majority of the 18,000 teachers could benefit from it. We have such a course at ASU and have observed that it is a great benefit to preservice teachers to mix them with experienced inservice teachers. Indeed, the course has been an eye-opener to regular physics majors who have stumbled into it, and we think that it would be valuable for graduate teaching assistants as well. Requiring at least one semester of the course for all physics majors would sensitize them to the profound challenges of physics education.

The "Methods course" should not stand alone; it should be an integral part of a balanced program for professional development. Its purpose is to fill a serious gap in the education of physics teachers. As asserted by the *National Standards* (1): "Effective science teaching is more than knowing science content and some teaching strategies. Skilled teachers of science have special understandings and abilities that integrate their knowledge of science content, curriculum, learning, teaching, and students. Such knowledge allows teachers to tailor learning situations to the needs of individuals and groups. This special knowledge, called "pedagogical content knowledge," distinguishes the science knowledge of teachers from that of scientists. It is one element that defines a professional teacher of science. In addition to solid knowledge of science, teachers of science must have a firm grounding in learning theory—understanding how learning occurs and is facilitated."

To integrate the Methods course into a professional development program the following conditions should be satisfied:
- peer teaching
- summer scheduling
- LPA follow-up.

These points require some elaboration.

The "*peer teaching principle*" holds that professionals are best taught by peers who are exceptionally well-versed in the objectives, methods and problems of the profession. Accordingly, the Methods course should be taught by a *master* inservice teacher. University faculty rarely have the pedagogical knowledge or the intimate familiarity with the high school scene to teach such a course. Moreover, the inservice teacher is a *stakeholder* in the teaching profession and therefore cares deeply about the course outcomes. It could be valuable, though not essential, for a physics professor to assist in teaching the course. That would enable faculty to get to the teachers personally, become sensitive to their needs and to contribute physics expertise. Some universities may require faculty participation to award academic credit for the course, but that is a local matter. Ideally, the master teacher would be given adjunct professor status and treated as a respected colleague by the faculty. This connection could be considerably strengthened by giving the master teacher a half time appointment during the academic year to work directly with physics teachers in the LPA.

The Methods course must be scheduled at a time when the teachers can attend. At an urban university within commuting distance, an evening or Saturday schedule would be feasible. However, summer scheduling is much to be preferred. From our experience, we know that a "workshop format" which immerses the teachers in collaborative efforts for the better part of each day for a minimum of four weeks is most effective, because it stimulates the formation of strong working relationships among the teachers. That is possible only in a summer course offered for up to six academic credits. By consultation with the LPA a sufficient enrollment of inservice teachers can be assured in advance. Moreover, teachers from rural schools can be included, especially if some provisions are made to help defray their expenses to attend.

The LPA, as a true learning community, should follow-up the Methods course with peer collaboration to help teachers implement and extend what they have learned. Teachers should be organized into *action research* teams which aim to solve practical problems to improve classroom teaching. As the *National Standards* assert (1): "Inquiry into practice is essential for effective teaching. Teachers need continuous opportunities to do so. Through collaborations with colleagues, teachers should inquire into their own practice." The LPA provides a forum for action research teams to share the results of their efforts with peers. Action research is a means for continuous critique and improvement of teaching practice. Because few teachers know how to conduct such research, one purpose of the Methods course should be to get them started. Thereafter, guidance and assistance by a master teacher and other peers is probably necessary for most teachers.

The primary responsibility for professional development lies with the teachers themselves. However, they cannot fulfill it effectively without resources supplied by a university or college. Any physics department with the will can easily meet the minimal commitment to support an LPA in a significant professional development program. Indeed, commitment from a single professor may be sufficient. All that is required of the professor is to schedule a Methods course taught by a master teacher. A good LPA can take care of the rest. It is to be hoped, though, that from a minimal commitment much more will grow.

## 5. Training Master Teachers—agents for reform

The most problematic factor in the formation of professional development programs is the enlistment of master teachers to conduct Workshop-courses and guide action research teams. There are many capable physics teachers with extensive teaching experience, but few are sufficiently well-informed and prepared to teach the kind of Methods course that is needed. They need some special training.

The AAPT has a leadership training program which certifies exceptionally qualified high school physics teachers as *Physics Teaching Resource Agents* (PTRA). The objective of this program is to prepare a cadre of teachers who are willing and able to assist others in improving their teaching. Universities can be a great help, especially by contributing to the creation and maintenance of an infrastructure through which the PTRA teachers can act effectively. The natural way is a marriage of local alliances with the PTRA. About 400 PTRAs have been trained so far.

PTRA training has focused on implementing technology, physics demos and some bits of physics education research. A systematic approach to physics pedagogy has been lacking. Our NSF-funded *Modeling Workshop* program aims to remedy this deficiency with intensive training over a three year period. (See (2) and references therein for details.) The Workshops in this program model in content and duration the kind of Methods course needed for local professional development programs. Out of more than 50 participants in the first phase of the program, we have identified 22 who are capable, eager and well-trained to function as master teachers. Many more will emerge from subsequent phases of the program. This program could be adapted and expanded into a national program for training and supporting master teachers nationwide.

## 6. Cooperation with schools and wider reform

As already described, university-LPA partnerships for professional development can be organized and operated without involving schools or school districts. In the long run, however, cooperation with schools will be necessary to improve teaching conditions. Most teachers have little time and meager resources for teaching innovation. The average physics class is allocated only $250 for supplies (3). Nevertheless, only 12% of physics teachers are dissatisfied with the

support they receive from the school administration (3). Indeed, in our work with teachers we have found that most schools are able to supply funds for computers, release time and other affordances when it is justified by the teacher's involvement in a program with the credibility of NSF support.

      A major reason that physics teachers have insufficient resources for innovation is that principals and school boards do not understand their needs or respond to unusual requests by individual teachers. Involvement in professional development partnerships can give teachers *leverage* for reforms in their schools through credible backing by universities. Strong university backing for a coherent professional development program can induce schools, school districts and communities to vie for its benefits. Thus, they may be induced to reduce contact hours, give release time for action research or pay teachers to participate in the summer workshops. In fact, many school districts and funding agencies have earmarked funds to support professional development, but they have trouble finding good ways to spend them. Great strides in science education reform could be made through regular support of teacher professional development in the summers. Adequate programs do not yet exist. However, the physics community could institute such a program nationwide in just a few years. The resources are in hand, only the will is needed.

      A strong nationwide professional development program for physics teachers would be a great help to the entire K-12 science education reform movement. High school physics is usually neglected in plans and programs for science education reform. Two reasons are commonly given: First, physics is regarded as elitist and suitable for only a small minority of students. Second, science teaching reform in primary and middle school is regarded as a more serious problem which must be addressed first. Let us see what is wrong with these views.

      We know that physics is an essential component of science and technology, so its place in science education at all levels must be strengthened. Between 20 and 25% of high school students take physics, but many more would do so if the teaching and course reputation were improved. Indeed, in schools with outstanding physics teachers the enrollment is invariably much higher than the norm. By improving the quality and relevance of physics courses, professional development partnerships could make this the rule rather than the exception. This is certainly the best way to increase physics enrollment — better than simply making physics a required course. If physics is to be taken by the majority of high school students, instruction must be reformed to make it worth their while.

      The bottom-up approach to science education reform focusing on primary and middle school teachers overlooks the power of a top-down approach beginning with high school physics at the top. Both approaches are needed, and the latter may be essential success of the former. Though professional development is widely recognized as essential to reform at all grade levels, many efforts are foundering for lack of qualified leaders—experienced teachers with strong scientific and pedagogical backgrounds. We have found that physics teachers with the pedagogical training supplied by our Modeling Workshops are

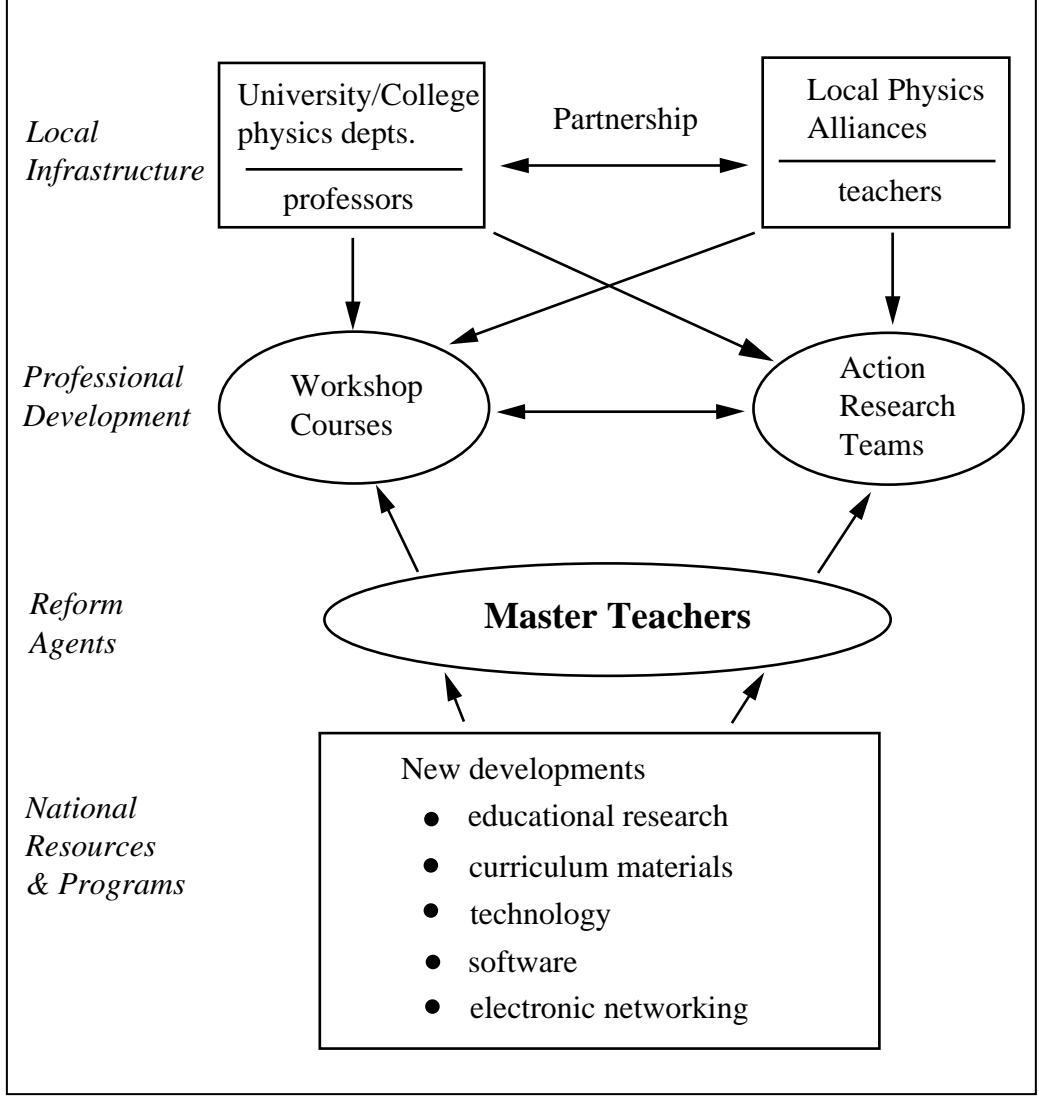

**FIGURE 2.**

well prepared and eager to help other science teachers at lower grade levels. Already, many of them are in demand for this purpose. We are thus confident in asserting that *a strong professional development for physics teachers will serve as a source of excellent inservice leaders for science education reform generally.*

# 7. A Blueprint for Sustained and Rapid Physics Teaching Reform

Physics education research and development has recently emerged as a respectable academic discipline within university physics departments. This will accelerate new developments in physics pedagogy, curriculum materials and technology-enabled instruction. We need a rapid delivery system to incorporate such advances into classroom practice and thereby drive educational reform. The traditional educational delivery system has enormous inertia and resistance to change. We need to short circuit it with direct delivery to inservice teachers. We cannot wait for the weighty action of academic curriculum committees, state certification boards, school boards and the like.

Figure 2 is the schematic for a rapid response system to drive physics education reform. It requires an infrastructure to keep a cadre of master teachers well-informed about new developments in physics education. The master teachers will be the principal agents of reform, linking educational research and development to the reform of classroom teaching. They will incorporate new results into their Workshop courses and feed new ideas and materials to action research teams to be evaluated and adapted for classroom use by the teachers. Thereby the master teachers will provide a continual stimulus to local partnerships for professional development.

Our Modeling Workshop Project (2) offers a ready framework for training and supporting master teachers. It provides a realization of the *National Standards* in a robust and flexible pedagogical framework for evaluating and integrating new pedagogical ideas and curriculum materials into effective instructional designs. It is thus an open system which can assimilate valuable instructional innovations from any quarter of the physics community for rapid dispersal to active teachers.

It is a system in which *inservice teachers are the final arbiters of pedagogical value.*

**Acknowledgment:** This work has been supported by a grant from the National Science Foundation.

# References

(1) National Research Council, *National Science Education Standards*, National Academy Press, Wash. DC, 1996.
(2) Hestenes, D., Modeling Methodology for Physics Teachers, These Proceedings.
(3) Neueschatz, M. & Alpert, L., 1993 Nationwide Survey of High School Teachers, American Institute of Physics, College Park, 1996.